\documentclass[epj,draft]{svjour}

\usepackage{latexsym}
\usepackage{graphics}

\begin{document}

\title{Nonequilibrium fluctuation induced escape from a metastable state}

\author{
J. Ray Chaudhuri\inst{1}\thanks{\email{jprc-8@yahoo.com}}, 
D. Barik\inst{2}\thanks{\email{pcdb5@mahendra.iacs.res.in}}
\and 
S. K. Banik\inst{3}\thanks{\email{skbanik@phys.vt.edu}}
}

\institute{
\inst{1} Department of Physics, Katwa College, Katwa,
Burdwan 713130, West Bengal, India. \\
\inst{2} Indian Association for the Cultivation of Science, Jadavpur, 
Kolkata 700032, India. \\
\inst{3} Department of Physics, Virginia Polytechnic Institute and
State University, Blacksburg, VA 24061-0435, USA.
}

\date{\today}

\abstract{
Based on a simple microscopic model where the bath is in a
non-equilibrium state we study the escape from a metastable state in
the over-damped limit. Making use of Fokker-Planck-Smoluchowski
description we derive the time dependent escape rate in the
non-stationary regime in closed analytical form which brings on to
fore a strong non-exponential kinetic of the system mode.
}

\PACS{
{05.40.-a}{Fluctuation phenomena, random processes, noise, and
Brownian motion} \and {02.50.Ey}{Stochastic processes}}

\maketitle

The problem of activated rate processes deals with the escape of a
Brownian particle from a metastable state under the influence of
thermal fluctuations generated by the immediate surrounding to which
the Brownian particle is in close contact. Based on nonequilibrium
statistical mechanics Kramers \cite{HAK} proposed a framework for
the phenomenon which over several decades became a standard paradigm
for theoretical and experimental investigation in many areas of
natural science \cite{HTB,VIM,AS,AHZ,LIM}. To the best of our
knowledge majority of the post Kramers developments of the theory
have been made in the stationary domain with few approaches in the
non-stationary regime within the framework of reactive flux
formalism \cite{HTB}. However, few attempts have been made to deal
the problem using nonequilibrium, non-stationary formalism
\cite{MMM,JRC,jayannavar}.

In their work Millonas and Ray \cite{MMM} proposed a theoretical
framework for studying the dynamics of escape rate from a metastable
state in the over-damped limit. Because of the nonequilibrium
fluctuations of the bath mode an in built fluctuating barrier
appears in the effective potential of the nonlinear Langevin
equation of the system variable. Using path integral formalism the
authors then derived a time dependent escape rate. Motivated by this
work \cite{MMM} we prescribe here an alternative method to derive
the time dependent escape rate using the Fokker-Planck-Smoluchowski
description. The object of the present work is twofold: First, to
consider a simple variant of system-bath model \cite{MMM} to study
the activated rate processes, where the associated heat bath is in
nonequilibrium state. The model incorporates some of the essential
features of Langevin dynamics with a fluctuating barrier which has
been phenomenologically proposed earlier \cite{MMM,PR}. Second,
since the theories of activated rate processes traditionally deal
with stationary bath, the non-stationary activated rate processes
have remained largely overlooked so far. We specifically address
this issue and examine the influence of initial excitation and
subsequent relaxation of bath modes on the activation of the
reaction coordinates within the framework of
Fokker-Planck-Smoluchowski equation. In spite of the fact that our
development bears a close kinship with the work of Millonas and Ray
\cite{MMM}, it is crucial to highlight that while Millonas and Ray
have used an explicit path integral approach towards the solution of
the problem, we, on the contrary, implement a naive differential
equation based approach which leads us to a {\em closed analytical
expression for the time-dependent escape rate}. We also mention that
in this work we have explicitly calculated the {\em non-exponential
kinetics} of the system mode, where the associated bath is not in
thermal equilibrium. The closed form of the final expression of our
approach brings with it the twin advantages of being capable of (1)
handling the non-stationary phenomena, and (2) tracing the
trajectory of how a system coupled with a non-equilibrium bath
reaches the stationary state in a computationally economic manner.

To make the present work self consistent we describe the essential
features of the model proposed by Millonas and Ray \cite{MMM}. The
physical situation that has been addressed is the following. At
$t=0_{-}$, the time just before the system and the bath are
subjected to an external excitations, the system is approximately
thermalized. At $t=0$, the excitation is switched on and the bath is
thrown into a non-stationary state which behaves as a nonequilibrium
bath. We follow the stochastic dynamics of the system mode after
$t>0$. The important separation of the time scale of the
fluctuations of the nonequilibrium bath and the thermal bath is that
the former effectively remains stationary on the fast correlation
time scale of the thermal noise.

The model consists of a system mode coupled to a set of relaxing
modes considered as a semi-infinite dimensional system
($\{q_k\}$-subsystem) which effectively constitutes a nonequilibrium
bath. This, in turn, is in contact with a thermally equilibrated
bath. Both the baths are composed of two sets of harmonic
oscillators characterized by the frequency sets $\{\omega_k\}$ and
$\{\Omega_j\}$ for the nonequilibrium and equilibrium baths,
respectively. The system-bath combination evolves under the total
Hamiltonian \cite{MMM}
\begin{eqnarray}\label{1.2}
H & = & \frac{p^2}{2 m}+V(x)+\frac{1}{2}\sum_j(P_j^2+\Omega_j^2Q_j^2)
+\frac{1}{2}\sum_k(p_k^2+\omega_k^2q_k^2)\nonumber \\
&& -x\sum_j\kappa_jQ_j
-g(x)\sum_k q_k-\sum_{j,k} \alpha_{jk}q_k Q_j .
\end{eqnarray}

\noindent The first two terms on the right hand side describe the
system mode. The Hamiltonian for the thermal and nonequilibrium
baths are described by the sets $\{Q_j,P_j\}$ and $\{q_j,p_j\}$ for
coordinate and momenta, respectively. The coupling terms containing
$\kappa_j$ refer to the usual system-bath linear coupling. The last
two terms indicate the coupling of the nonequilibrium bath to the
system and the thermal bath modes, respectively. In the present
problem $H$ is considered to be classical and the temperature $T$ is
high for the thermally activated problem, so that the quantum
effects do not play any significant role. For simplicity we take
mass, $m=1$ in equation (\ref{1.2}) and for the rest of the
treatment. The form of the nonequilibrium bath, a set of phonons, is
chosen for both simplicity and because of its generic relationship
to many condensed matter systems.

Elimination of equilibrium reservoir variables $\{Q_j,P_j\}$ in an
appropriate way we have the equation of motion for the
nonequilibrium bath modes as \cite{JRC,MID,KL}
\begin{equation}\label{1.3}
\ddot{q}_k+\gamma\dot{q}_k+\omega_k^2q_k=g(x)+\eta_k(t) .
\end{equation}

\noindent This takes into account the average dissipation ($\gamma$)
of the nonequilibrium bath modes $q_k$ due to their coupling to the
thermal bath which induces fluctuations $\eta_k(t)$ characterized by
$\langle \eta_k (t) \rangle=0$ and $\langle \eta_j(t)\eta_k(t')
\rangle=2\gamma k_BT\delta(t-t')\delta_{jk}$. In moving from
equation (\ref{1.2}) to equation (\ref{1.3}) the cross terms of the
form $\sum_{j}\gamma_{kj}q_j$ has been neglected for $j\ne k$.
Proceeding similarly to eliminate the thermal bath variables from
the equation of motion of the system mode, we get
\begin{equation}\label{1.4}
\ddot{x}+\gamma_{eq}\dot{x}+V'(x)=\xi_{eq}(t)+g'(x)\sum_k q_k ,
\end{equation}

\noindent
where $\gamma_{eq}$ refers to the dissipation coefficient of the
system mode due to its coupling to the thermal bath providing fluctuations
$\xi_{eq}(t)$ with the properties,
\begin{equation}
\langle \xi_{eq} (t) \rangle=0, 
\langle \xi_{eq}(t)\xi_{eq}(t') \rangle=2\gamma_{eq} k_BT\delta(t-t').
\end{equation}

Now making use of the formal solution of equation (\ref{1.3})
which takes into account of the relaxation of the nonequilibrium modes,
and integrating over the nonequilibrium modes with a Debye type
frequency distribution of the form
\begin{eqnarray}\nonumber
\cal{D}(\omega)&=&3\omega^2/2\omega_c^3 \; \; for \; \; |\omega|\le\omega_c\\
&=& 0  \; \; for \; \; |\omega|>\omega_c \nonumber
\end{eqnarray}

\noindent
where $\omega_c$ is the high frequency Debye-cut-off, we finally
arrive at the following Langevin equation of motion for the system
mode,
\begin{equation}\label{1.5}
\ddot{x}+\Gamma(x)\dot{x}+\tilde{V}'(x)=\xi_{eq}(t)+g'(x)\xi_{neq}(t).
\end{equation}

\noindent
Here $\Gamma(x)$ is a system coordinate dependent dissipation constant
and is given by
\begin{equation}\label{1.6}
\Gamma(x)=\gamma_{eq}+\gamma_{neq}[g'(x)]^2
\end{equation}

\noindent
and $\xi_{neq}(t)$ refers to the fluctuations of the nonequilibrium bath
modes which effectively cause a damping of the system mode by an
amount $\gamma_{neq}[g'(x)]^2$. Equation (\ref{1.5}) also includes
the modification of the bare potential $V(x)$
\begin{equation}\label{1.7}
\tilde{V}(x)=V(x)-\frac{\omega_c}{\pi}\gamma_{neq}g^2(x) .
\end{equation}

\noindent equation (\ref{1.5}) thus describes the effective dynamics
of a particle in a modified barrier, where the metastability of the
well originates from the dynamic coupling $g(x)$ of the system mode
with the nonequilibrium bath modes.

In order to define the dynamics described by equation (\ref{1.5})
completely it is necessary to state the properties of the
fluctuations of the nonequilibrium bath $\xi_{neq}(t)$, which is
assumed to be Gaussian with zero mean $\langle
\xi_{neq}(t)\rangle=0$. Also the essential properties of
$\xi_{neq}(t)$ explicitly depend on the nonequilibrium state of the
intermediate oscillator modes $\{q_k\}$ through $u(\omega, t)$, the
energy density distribution function at time $t$ in terms of the
fluctuation-dissipation relation for the nonequilibrium bath
\cite{MMM}
\begin{eqnarray}
u(\omega, t)&=&\frac{1}{4\gamma_{neq}}\int_{-\infty}^{+\infty}
d\tau \langle \xi_{neq}(t)\xi_{neq}(t+\tau) \rangle e^{i\omega \tau}\nonumber\\
&=&\frac{1}{2}k_BT+e^{-\gamma t/2}\left[u(\omega, 0)-
\frac{1}{2}k_BT\right] ,\label{1.8}
\end{eqnarray}

\noindent where $\left[u(\omega, 0)-\frac{1}{2}k_BT\right]$ is a
measure of departure of energy density from thermal average at
$t=0$. The exponential term implies that deviation due to the
initial excitation decays asymptotically to zero as $t\rightarrow
\infty$, so that one recovers the usual fluctuation-dissipation
relation for the thermal bath. With the above specification of
correlation function of $\xi_{neq}$, equation (\ref{1.8}) thus
attributes the non-stationary character of $\{q_k\}$-subsystem.

On time scales larger than the inverse friction coefficient
$1/\gamma_{eq}$, we can in most particular cases consider the
over-damped limit of the Langevin equation. This in turn corresponds
to the adiabatic elimination of the fast variables, inertia term,
from the equation of motion by putting $\ddot{x}=0$ for homogeneous
systems. In contrast, for the case of inhomogeneous system the above
method of elimination does not work properly and Sancho \textit{et
al} \cite{JMS} have given a proper prescription for the elimination
of fast variables. Using the method of Sancho \textit{et al} the
formal master equation for the probability density of the process
$P(x,t)=\langle \rho(x,t) \rangle$ is given by
\begin{eqnarray}
\frac{\partial P}{\partial t}&=&\frac{\partial}{\partial x}\left\{
\frac{\tilde{V}'(x)}{\Gamma(x)}P \right\}+\gamma_{eq}k_BT\frac{\partial}
{\partial x}\left\{\frac{1}{\Gamma(x)}\frac{\partial}{\partial x}
\frac{1}{\Gamma(x)}P \right\}\nonumber\\
&& +\gamma_{neq}k_BT\left(1+r e^{-\gamma t/2}\right)\frac{\partial}{\partial x}
\left\{ \frac{g'(x)}{\Gamma(x)}\frac{\partial}{\partial x}
\frac{g'(x)}{\Gamma(x)}P \right\} \nonumber \\
&& + \gamma_{neq}k_BT\left(1+r e^{-\gamma
t/2}\right)\frac{\partial}{\partial x}
\left\{\frac{g'(x)g''(x)}{\Gamma^2(x)} P\right\} \label{1.9}
\end{eqnarray}

\noindent where $r=\{[u(\omega\rightarrow 0, 0)/2k_BT]-1\}$ is a
measure of the deviation from equilibrium at the initial instant.
equation (\ref{1.9}) is the Fokker-Planck-Smoluchowski equation
where the associated bath is in nonequilibrium state, and is the
\textit{first key result of this paper}. Under stationary condition
(at $t\rightarrow \infty$) $\partial P/\partial t=0$ and the
stationary distribution obeys the equation
\begin{equation}\label{1.10}
k_BT\frac{dP_{st}(x)}{dx}+\tilde{V}'(x)P_{st}(x)=0
\end{equation}

\noindent
which has the solution
\begin{equation}\label{1.11}
P_{st}(x)=N\exp\left[ -\frac{1}{k_BT}\int^x \tilde{V}'(x') dx'
\right]
\end{equation}

\noindent where $N$ is the normalization constant. In ordinary
Strato- novich description the Langevin equation corresponding to the
Fokker-Planck-Smoluchowski equation (\ref{1.9}) is given by
\begin{equation}\label{1.12}
\dot{x}=-\frac{\tilde{V}'(x)}{\Gamma(x)}-\frac{D(t)g''(x)g'(x)}{\Gamma^2(x)}
+\frac{1}{\Gamma(x)}\xi_{eq}(t)+\frac{g'(x)}{\Gamma(x)}\xi_{neq}(t)
\end{equation}

\noindent
where
\begin{equation}\label{1.13}
D(t)=\gamma_{neq}k_BT(1+\gamma e^{-\gamma t/2})
\end{equation}

\noindent is the time-dependent diffusion constant due to the
relaxation of nonequilibrium bath. Using equations (\ref{1.9}) or
(\ref{1.12}) the escape rate from a metastable state can be
calculated via steepest descent method \cite{HR}
\begin{equation}\label{1.14}
k=\frac{\tilde{\omega}_0 \tilde{\omega}_b}{2 \pi
\Gamma(x_b)}\exp\left(-\frac{\tilde{E}_b}{k_BT}\right)
\end{equation}

\noindent where $k$ is the Kramers activation rate, with
$\tilde{E}_b=\tilde{V}'(x_b)-\tilde{V}'(x_0)$ is the modified
activation energy and $\tilde{\omega}_b=[\tilde{V}''(x_b)]^{1/2}$,
$\tilde{\omega}_0=[\tilde{V}''(x_0)]^{1/2}$ are the modified
frequencies at the barrier top and the bottom of the potential well,
respectively. $x_b$ denotes the position of the barrier top and
$x_0$ is the position of the bottom of the potential well. In
equation (\ref{1.14}) $\Gamma$ has been evaluated at the top of the
barrier. In the absence of the nonequilibrium bath (\ref{1.14})
reduces to standard Kramers' expression \cite{HAK},
\begin{equation}
k=\frac{\omega_0 \omega_b}{2 \pi\gamma}
\exp\left(-\frac{E_b}{k_BT}\right).
\end{equation}

To obtain the time dependent rate $k(t)$, let us consider that the
time dependent solution of equation (\ref{1.9}) is given by
\begin{equation}\label{1.15}
P(x,t)=P_{st}(x)e^{-\phi(t)}
\end{equation}

\noindent
where $\phi$ is a function of $t$ only and $\lim_{t \rightarrow
\infty}\phi(t)=0$. $P_{st}(x)$ is the steady state solution of
equation (\ref{1.9}). Substitution of (\ref{1.15}) in (\ref{1.9})
separates the space and time parts and we have the equation for
$\phi(t)$ as
\begin{equation}\label{1.16}
-\frac{d\phi}{dt}e^{\gamma
t/2}=constant=\alpha (say)
\end{equation}

\noindent
which after integration over time gives
\begin{equation}\label{1.17}
\phi(t)=\frac{2\alpha}{\gamma}e^{-\gamma t/2}
\end{equation}

\noindent where $\alpha$ can be determined by initial condition. The
time dependent solution of equation (\ref{1.9}) therefore reads
\begin{equation}\label{1.18}
P(x,t)=P_{st}(x)\exp\left[ -\frac{2 \alpha}{\gamma}e^{-\gamma
t/2}\right].
\end{equation}

\noindent To determine $\alpha$ we now assume that \cite{JRC} just
at the moment the system (and the non-thermal bath) is subjected to
external excitation at $t=0$, the distribution must coincide with
the usual Boltzmann distribution where the energy term in the
Boltzmann factor in addition to the usual kinetic and potential
energy terms, contains the initial fluctuations of energy density
$\Delta u[=u(\omega, 0)-\frac{1}{2}k_BT]$ due to excitation of the
system at $t=0$. This gives $\alpha=(\gamma/2)(\Delta u/k_BT)$.
$\alpha$ is thus determined in terms of relaxing mode parameters and
fluctuations of the energy density distribution at $t=0$. The time
dependent rate is then derived as \cite{HR}
\begin{equation}\label{1.19}
k(t)=k\exp\left[ -\frac{\Delta u}{k_BT}
e^{-\frac{\gamma}{2}t}\right]
\end{equation}

\noindent where $\Delta u$ is the measure of the initial departure
from the average energy density distribution due to the preparation
of the non-stationary state of the intermediate bath modes as a
result of excitation at $t=0$ and $k$ is given by equation
(\ref{1.14}). The above result, equation (\ref{1.19}), \textit{which
is the second key result of this paper}, illustrates a strong
nonexponential kinetic of the system mode undergoing a
non-stationary activated rate processes in the over-damped regime.
The origin of this is an initial preparation of nonequilibrium mode
density distribution which eventually relaxes to an equilibrium
distribution. equation (\ref{1.19}) implies that the initial
transient rate is different from the asymptotic steady state
Kramers' rate. The sign of $\Delta u$ determines whether the initial
rate will be faster or slower than the steady state rate. This is
because there exists a time lag for the non-thermal energy gained by
the few nonequilibrium modes by sudden excitation to be distributed
over a range before it becomes available to the reaction coordinate
as thermal energy for activation.

In conclusion, based on a system-reservoir model, where the
reservoir is in a non-equilibrium state, we have provided an
analytic model to derive the closed time-dependent escape rate from
a metastable state induced by non-equilib- rium fluctuations. We have
explicitly calculated the {\em non-exponential kinetics} of the
system mode, where the associated bath is not in thermal
equilibrium. Our methodology takes care of the non-stationary
phenomena, and simultaneously traces the barrier dynamics of a
system when it is coupled with a non-equilibrium bath. Not only that
our approach may serve as a potential avenue towards the explanation
of non-stationary transport processes and rachet problems envisaged
in various chemically and biologically interesting systems. The work
in this direction is in progress in our group.

\begin{acknowledgement}
This work is partially supported by Indian Academy of Sciences,
Bangalore. SKB acknowledges support from Virginia Tech through
ASPIRES award program.
\end{acknowledgement}

\end{document}